\documentclass[manuscript]{aastex}

\usepackage{mathbbol}
\usepackage{amsmath}

\shorttitle{Turbulence transport and application to CMEs}
\shortauthors{Wiengarten et al.}

\begin{document}

\title{Implementing turbulence transport in the CRONOS framework and application to the propagation of CMEs}


\author{T. Wiengarten\altaffilmark{1}, H. Fichtner\altaffilmark{1} and J. Kleimann\altaffilmark{1}}
\affil{\altaffilmark{1}Institut f\"ur Theoretische Physik IV, Ruhr-Universit\"at Bochum, Germany}

\and

\author{R. Kissmann\altaffilmark{2}}
\affil{\altaffilmark{2}Institut f\"ur Astro- und Teilchenphysik, Universit\"at Innsbruck, Austria}




\begin{abstract}
We present the implementation of turbulence transport equations in addition to the Reynolds-averaged MHD equations within the {\sc Cronos} framework. 
The model is validated by comparisons with earlier findings before it is extended to be applicable to regions
in the solar wind that are not highly super-Alfv\'enic. We find that the respective additional terms result in absolute normalized cross-helicity to decline more slowly, while a proper implementation of the mixing terms can even lead to increased cross-helicities in the inner heliosphere.\\
The model extension allows to place the inner boundary of the simulations closer to the Sun, where we choose its location at 0.1~AU for future application to the Wang-Sheeley-Arge model. Here, we concentrate on effects on the turbulence evolution for transient events by injecting a coronal mass
ejection (CME). We find that the steep gradients and shocks associated with these structures result in enhanced turbulence levels and reduced
cross-helicity. Our results can now be used straightforwardly for studying the transport of charged energetic particles, where the elements of the diffusion tensor can now benefit from the self-consistently computed solar wind turbulence. Furthermore, we find that there is no strong back-reaction of the turbulence on the large-scale flow so that CME studies concentrating on the latter need not be
extended to include turbulence transport effects. 
\end{abstract}



\section{Introduction}
In a recent publication \citep{Wiengarten-etal-2014} we presented results for inner-heliospheric solar wind conditions from simulations
based on observational boundary conditions derived from the Wang-Sheeley-Arge (WSA, \citet{Arge-Pizzo-2000}) model. The obtained 3D 
configurations provide the basis for subsequent studies of energetic particle transport. 
For a long time, the standard approach to couple the solar wind dynamics to the transport of charged energetic particles
has been to parameterise all transport processes in terms of `background' quantities such as
the large-scale solar wind velocity and heliospheric magnetic field strength. This was particularly true 
for the treatment of spatial diffusion \citep[see, e.g.,][]{Potgieter-2013}. Only in recent years so-called 
{\it ab-initio} models were developed, in which the diffusion coefficients are formulated explicitly as
functions of small-scale, low-frequency turbulence quantities such as the magnetic field fluctuation amplitude
or the correlation length \citep[e.g.,][]{Parhi-etal-2003, Pei-etal-2010, Engelbrecht-Burger-2013}. This
significant improvement has been possible, on the one hand due to the development of turbulence transport
models that, most often, describe the evolution of turbulent energy density, cross-helicity, and correlation 
lengths for low-frequency turbulence, as observed in the solar wind, with increasing distance from the Sun,
see, e.g., the review in \citet{Zank-2014}. On the other hand, our present-day knowledge about turbulence
in the solar wind \citep[see, e.g., the reviews by][]{Matthaeus-Velli-2011,Bruno-Carbone-2013} has increased
tremendously since the first measurements \citep{Coleman-1968, Belcher-Davis-1971} thanks to highly
sophisticated analyses \citep[e.g.,][]{Horbury-etal-2008,Horbury-etal-2012}. 

After the pioneering papers by \citet{Tu-etal-1984} and \citet{Zhou-Matthaeus-1990a,Zhou-Matthaeus-1990b,Zhou-Matthaeus-1990c} and first systematic 
studies of the radial evolution of turbulence quantities \citep{Zank-etal-1996}, major progress has only 
been achieved in recent years. First, \citet{Breech-etal-2008} improved the previous modeling by
considering off-ecliptic latitudes. Second, in \citet{Usmanov-etal-2011} this turbulence model was 
extended to full time dependence and three spatial dimensions. In the same paper the turbulence evolution 
equations were simultaneously solved along with a large-scale MHD model of the supersonic solar wind.
While \citet{Usmanov-etal-2012, Usmanov-etal-2014} extended this study further with the incorporation of
pick-up ions and electrons as separate fluids, as well as an eddy-viscosity approximation to self-consistently account for turbulence driven by shear,
\citet{Oughton-etal-2011} undertook another extension of 
the modeling by considering not only one but two, mutually interacting, incompressible components, namely
quasi-two-dimensional turbulent and wave-like fluctuations. The effect on turbulence due to changing solar wind conditions in the outer heliosphere during the solar cycle have recently been addressed by \citet{Adhikari-etal-2014}.
   
All of the mentioned models have one limitation in common, namely the condition that the Alfv\'en speed $V_A$ is
significantly lower than that of the solar wind $U$, which precludes their application to the heliocentric distance range
below about 0.3~AU. The turbulence in this region and its consequences for the dynamics of the solar wind 
have been studied on the basis of the somewhat simplified transport equations for the wave power spectrum 
and the wave pressure \citep[e.g.,][]{Tu-Marsch-1995, Hu-etal-1999, Vainio-etal-2003, 
Shergelashvili-Fichtner-2012}. The Alfv\'en speed limitation in the more detailed models was very
recently removed with a new, comprehensive approach to the general problem of the transport of low-frequency
turbulence in astrophysical flows by \citet{Zank-etal-2012a}, see also \citet{Dosch-etal-2013}. Not only are 
the derived equations formally valid close to the Sun in the sub-Alfv\'enic regime of the solar wind, but they also
represent an extension of the treatment of turbulence by non-parametrically and quantitatively considering
the evolution of the so-called energy difference in velocity and magnetic field fluctuations, and by 
explicitly describing correlation lengths for sunward and anti-sunward propagating fluctuations as well 
as for the energy difference (sometimes also referred to as the residual energy). 

In our previous modeling \citep{Wiengarten-etal-2014} we derived inner boundary conditions beyond the Alfv\'enic critical point at 0.1~AU by means of the WSA model. This approach matches a potential field solution for the coronal magnetic field to the observed photospheric magnetograms, and the resulting magnetic field topology can be linked empirically to the corresponding solar wind speed for every open field line \citep{Wang-Sheeley-1990}. We performed simulations of the inner-heliospheric solar wind conditions for several Carrington rotations. Subsequently, the resulting 
time-dependent 3D configuration is used to study the transport of charged energetic particles by means of stochastic differential equations (A.~Kopp, priv.~comm.). 
Within that study, the transport coefficients required for the latter model have so far been estimated from the large-scale quantities provided by the ideal MHD equations (see discussion above). 
In order to obtain more realistic transport coefficients for the study of propagation of charged energetic particles, we extend our
previous modeling by solving the 3D time-dependent turbulence tranport equations self-consistently coupled to the Reynolds-averaged
MHD equations. Such an approach was taken by \citet{Usmanov-etal-2011} and will provide the starting point for our modeling. In a first step we validate our 
implementation by comparing with these authors' findings. The turbulence transport equations used there can also be obtained from the more
general turbulence transport equations of \citet{Zank-etal-2012a} by applying respective simplifications. The Usmanov model neglects the 
Alfv\'en velocity and is therefore only applicable to highly-super-Alfv\'enic solar wind regimes and we remove this constraint by retaining
the respective terms of the more general Zank model. This extention to the model
allows us to place the inner boundary of the simulations closer to the Sun, for which we choose 0.1~AU for future couplings
to the WSA model. A correction of the model is also required with regard to the unappropriate absence of 
turbulence driving by shear, which has only recently been addressed by \citet{Usmanov-etal-2014}. Here, we will follow 
earlier simple ad-hoc approaches as in \citet{Zank-etal-1996} and \citet{Breech-etal-2008}. This improvement is essential for the application to 
solar wind transients, such as coronal mass ejections (CMEs), which we address in the second part of this paper.

With this application to the propagation of coronal mass ejections (CMEs) we extend earlier
approaches towards a self-consistent treatment of the expansion of the disturbed solar 
wind. Such a treatment is desirable because it has been realized that a proper modeling of
CMEs requires a realistic model for the `background' solar wind 
\citep[e.g.,][]{Jacobs-Poedts-2011, Lee-etal-2013}. If one does not resort 
to empirical solar wind background models \citep[like][]{Cohen-etal-2007},
one needs to model the solar wind turbulence explicitly \citep{Sokolov-etal-2013}. 
Besides the importance of a self-consistent incorporation of the turbulence evolution
in the solar wind for CME propagation studies \citep[see also][]{vanderHolst-etal-2014}, the 
generation of turbulence by large-scale disturbances is of interest for energetic
particle transport studies at interplanetary shocks \citep[e.g.,][]{Sokolov-etal-2009}. While a self-consistent 
turbulence incorporation into a model of the supersonic solar wind out to 100~AU
including corotating interaction regions has been achieved by \citet{Usmanov-etal-2012},
corresponding studies for the case of CMEs are still limited to the most inner heliosphere
inside about 30 solar radii \citep{Jin-etal-2013}. These studies use, compared to the approaches 
developed by \citet{Oughton-etal-2011}, \citet{Usmanov-etal-2012}, or 
\citet{Zank-etal-2012a}, simplified equations for the treatment of the turbulence evolution.
We extend the modeling here by both employing a refined treatment of the turbulence 
evolution and solving the self-consistent model equations out to 1.2~AU.

The paper is organized as follows: In Section \ref{sec:code} we present the turbulence transport equations and the respective coupling
to the ideal MHD equations. Details are outsourced to the Appendix. Section \ref{sec:validation} is used to validate our implementation
by comparing with the respective results of \citet{Usmanov-etal-2011}. We present the extensions we apply to the model and demonstrate
their effects in Section \ref{sec:extension}. In Section \ref{sec:innerheliosphere} we move the inner boundary closer to the Sun and show results for both quiet and CME-disturbed cases. 
We conclude with a summary and an outlook on future improvements in Section \ref{sec:conclusions}.


\section{Equations and Code Setup}
\label{sec:code}
We follow the approach of \citet{Usmanov-etal-2011} to incorporate the evolution equations of small-scale turbulence quantities into the framework of the MHD code {\sc Cronos} (see Appendix \ref{app_cronos} for details concerning the code). As we will seek to inject coronal mass ejections at an inner boundary of 0.1 AU, the assumption of highly-super-Alfv\'{e}nic solar wind conditions is not justifiable everywhere. The turbulence transport equations are extended to keep terms associated with the Alfv\'{e}n velocity, which can be derived from a simplified model by \citet{Zank-etal-2012a} (see Appendix \ref{appendixA0}), while we employ the coupling between small-scale and large-scale equations as given in \citet{Usmanov-etal-2011}. The turbulence transport equations for the considered case then are :
\begin{align}
\partial_t Z^2 + \nabla\cdot({\bf U}Z^2 + {\bf V}_AZ^2\sigma_C) &= \frac{Z^2(1-\sigma_D)}{2}\nabla\cdot{\bf U} + 2{\bf V}_A\cdot\nabla(Z^2\sigma_C) + Z^2\sigma_D{\bf\hat{B}}\cdot({\bf\hat{B}}\cdot\nabla){\bf U} \nonumber \\
& \ - \frac{\alpha Z^3f^+(\sigma_C)}{\lambda} + \langle{\bf z^+}\cdot{\bf S^+}\rangle +  \langle{\bf z^-}\cdot{\bf S^-}\rangle \label{eq:Z2} \\
\partial_t(Z^2\sigma_C) + \nabla\cdot({\bf U}Z^2\sigma_C + {\bf V}_AZ^2) &= \frac{Z^2\sigma_C}{2}\nabla\cdot{\bf U} + 2{\bf V}_A\cdot\nabla Z^2 + Z^2\sigma_D\nabla\cdot{\bf V}_A \nonumber \\ 
& \ - \frac{\alpha Z^3f^-(\sigma_C)}{\lambda} + \langle{\bf z^+}\cdot{\bf S^+}\rangle - \langle{\bf z^-}\cdot{\bf S^-}\rangle \label{eq:Z2sigC}\\
\partial_t(\rho\lambda) + \nabla\cdot({\bf U}\rho\lambda) &= \rho\beta\left[Zf^+(\sigma_C) - \frac{\lambda}{\alpha Z^2}\left(\langle{\bf z^+}\cdot{\bf S^+}\rangle(1-\sigma_C) + \langle{\bf z^-}\cdot{\bf S^-}\rangle(1+\sigma_C) \right)\right] \nonumber \\ \label{eq:lam}
\end{align}
with $f^\pm := \sqrt{1-\sigma_C^2}\left[\sqrt{1+\sigma_C}\pm\sqrt{1-\sigma_C}\right]$
where the following moments of the Els\"asser variables (described below) are used:\\
\begin{align}
Z^2 &:= \frac{\langle\bf{z^+}\cdot\bf{z^+}\rangle +  \langle\bf{z^-}\cdot\bf{z^-}\rangle}{2} = \langle u^2 \rangle + \langle b^2/\rho \rangle \\
Z^2\sigma_C &:= \frac{\langle\bf{z^+}\cdot\bf{z^+}\rangle -  \langle\bf{z^-}\cdot\bf{z^-}\rangle}{2} = 2\langle\bf{u}\cdot\bf{b/\sqrt{\rho}}\rangle \\
Z^2\sigma_D &:= \langle{\bf z^+}\cdot{\bf z^-}\rangle = \langle u^2 \rangle - \langle b^2/\rho \rangle ~.
\end{align}
Here, $Z^2$ is twice the total energy per unit mass of the fluctuations, $\sigma_C$ is the normalized cross-helicity, $\sigma_D$ is the normalized difference between the magnetic and kinetic energy of the fluctuations per unit mass, also known as residual energy. As done in previous studies, we assume an observationally inferred constant value for the energy difference $\sigma_D=-1/3$ \citep{Tu-Marsch-1995} and reserve an extension to include a variable energy difference \citep{Zank-etal-2012a} for future studies. Furthermore, $\lambda$ is the correlation length, ${\bf V}_A={\bf B}/\sqrt{\rho}$ is the normalized Alfv\'{e}n velocity, $\alpha=2\beta=0.8$ are the Karman-Taylor constants, and terms involving sources of turbulence ${\bf S}^\pm$ are discussed in subsequent sections.\\
The Els\"asser variables $\bf{z^\pm} := \bf{u} \pm \bf{b}/\sqrt{\rho}$, where $\bf{u}$ and $\bf{b}$ denote the fluctuations about the mean fields $\bf{U}$ and $\bf{B}$, describe the inward ($\bf{z^+}$) and outward ($\bf{z^-}$) propagating modes with respect to the mean magnetic field so that $\bf{z^\pm}$ is anti-parallel/parallel to it. $\bf{\hat{B}}$ denotes the unit vector in the direction of the mean magnetic field.\\   

The turbulence transport equations (\ref{eq:Z2}) -- (\ref{eq:lam}) are implemented in the framework of the {\sc Cronos} code as additional equations to be solved alongside the Reynolds-averaged, normalized ideal MHD equations in the co-rotating frame of reference with respective coupling terms to account for the effects of turbulence on the large-scale MHD quantities in analogy to \citet{Usmanov-etal-2011}:
\begin{align}
  \partial_t \rho &+ \nabla \cdot (\rho {\bf V}) = 0  \label{eq:conti}  \\
  \partial_t (\rho {\bf U}) &+ \nabla \cdot \left[\rho {\bf V U} 
    +  \bar{p} \ {\bf 1} - {\eta\bf B B} \right]
  = -\rho\left({\bf g}+{\bf\Omega}\times{\bf U}\right) \label{eq:mom}\\
	\partial_t {\bf B} &+ \nabla \cdot ({\bf VB-BV}) = 0   \label{eq:indu} \\
  \partial_t e &+ \nabla \cdot \left[ e{\bf V} + (p+ |{\bf B}|^2/2) \ {\bf U}
    - ({\bf U} \cdot {\bf B}) {\bf B} - {\bf V}_A\rho Z^2\sigma_C/2 + {\bf q}_H \right] = \nonumber \\
		&-\rho{\bf V}\cdot{\bf g} -{\bf U}\cdot\nabla p_w - \frac{Z^2\sigma_C}{2}{\bf V}_A\cdot\nabla\rho + \frac{\rho Z^3f^+(\sigma_C)}{2\lambda}\nonumber \\
	&+ {\bf U}\cdot({\bf B}\cdot\nabla)[(\eta-1){\bf B}] - \rho{\bf V}_A\cdot\nabla(Z^2\sigma_C)  
	\label{eq:energy}
\end{align}
with $\bar{p} = (p + |{\bf B}|^2/2 + p_w)$, $p_w = (\sigma_D+1)\rho Z^2/4$ and $\eta = 1+\sigma_D\rho Z^2/(2B^2)$ for the considered case of transverse and axisymmetric turbulence. Equations (\ref{eq:conti}) -- (\ref{eq:indu}) are identical to those of \citet{Usmanov-etal-2011}, while the above form of the energy equation is derived in Appendix \ref{appendixB}. Furthermore, $\rho$ is the mass density, ${\bf U}$ and ${\bf V} = {\bf U}-{\bf \Omega}\times{\bf r}$ denote the fluid velocity in the rest and corotating frame, respectively, ${\bf B}$ is the mean magnetic field, $p$ is the scalar thermal pressure, ${\bf g} = (GM_\odot/r^2){\bf\hat{r}}$ describes the Sun's gravitational acceleration, and ${\bf \Omega} = \Omega {\bf e}_z$ is the Sun's angular rotation speed with $\Omega = 14.71^\circ/{\rm d}$ \citep{Snodgrass-Ulrich-1990}. The energy density $e=\rho U^2/2 + B^2/2 + p/(\gamma-1)$ is used without both the gravitational potential and the turbulent energy component $\rho Z^2/2$, which are instead attributed by means of the right-hand sided source terms in Equation (\ref{eq:energy}), see Appendix \ref{appendixB}. An adiabatic equation of state is used with $\gamma=5/3$, while, due to the inclusion of Hollweg's heat flux \citep{Hollweg-1974, Hollweg-1976} ${\bf q}_H = (3/4)p{\bf V}$, the effective value of the adiabatic index $\gamma_{eff} = 13/9$ is close to observationally inferred values \citep{Totten-etal-1995}.  \\
We use  spherical coordinates ($r$,$\vartheta$,$\varphi$) with the origin being located at the center of the Sun. Thus, $r$ is the heliocentric radial distance, $\vartheta\in[0,\pi]$ is the colatitude or polar angle (with the north pole corresponding to $\vartheta=0$) and $\varphi\in[0,2\pi]$ is the azimuthal angle.\\
The above sets of equations are both given in their normalized form, so that for instance no factors of $4\pi$ or $\mu_0$ occur with the magnetic energy density.\\
{\sc Cronos} employs a semi-discrete finite-volume scheme with Runge-Kutta time integration and adaptive time-stepping, allowing for different approximate Riemann solvers. Although we make use of its conservative features by implementing as many terms as possible as divergence of fluxes, a number of source terms remain. In cases where source terms involve differentiation we apply second order accurate central finite differences. The solenoidality of the magnetic field is ensured via constrained transport, provided the magnetic field is initialized as divergence-free. Besides the code's support of Cartesian, cylindrical, and spherical (including coordinate singularities) coordinates, it also allows for non-equidistantly spaced grids, e.g. a spatially varying $\Delta r$, as long as all coordinate planes remain orthongonal.

\section{Model Validation}
\label{sec:validation}
\begin{figure}
\includegraphics[width=\textwidth]{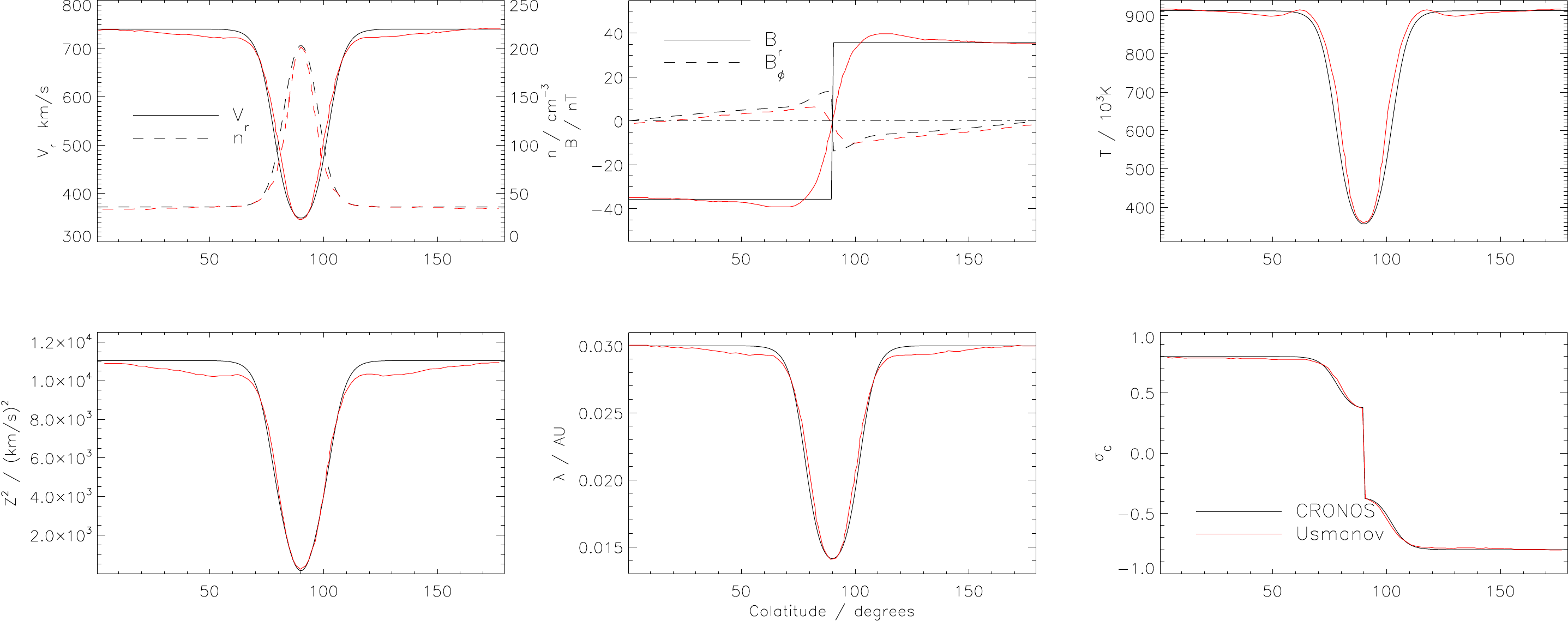}
\caption[]{Boundary conditions at 0.3 AU (black lines) in comparison with those of \citet{Usmanov-etal-2011} (red lines).}
\label{fig:Usm-init}
\end{figure}

\begin{figure}
\includegraphics[width=\textwidth]{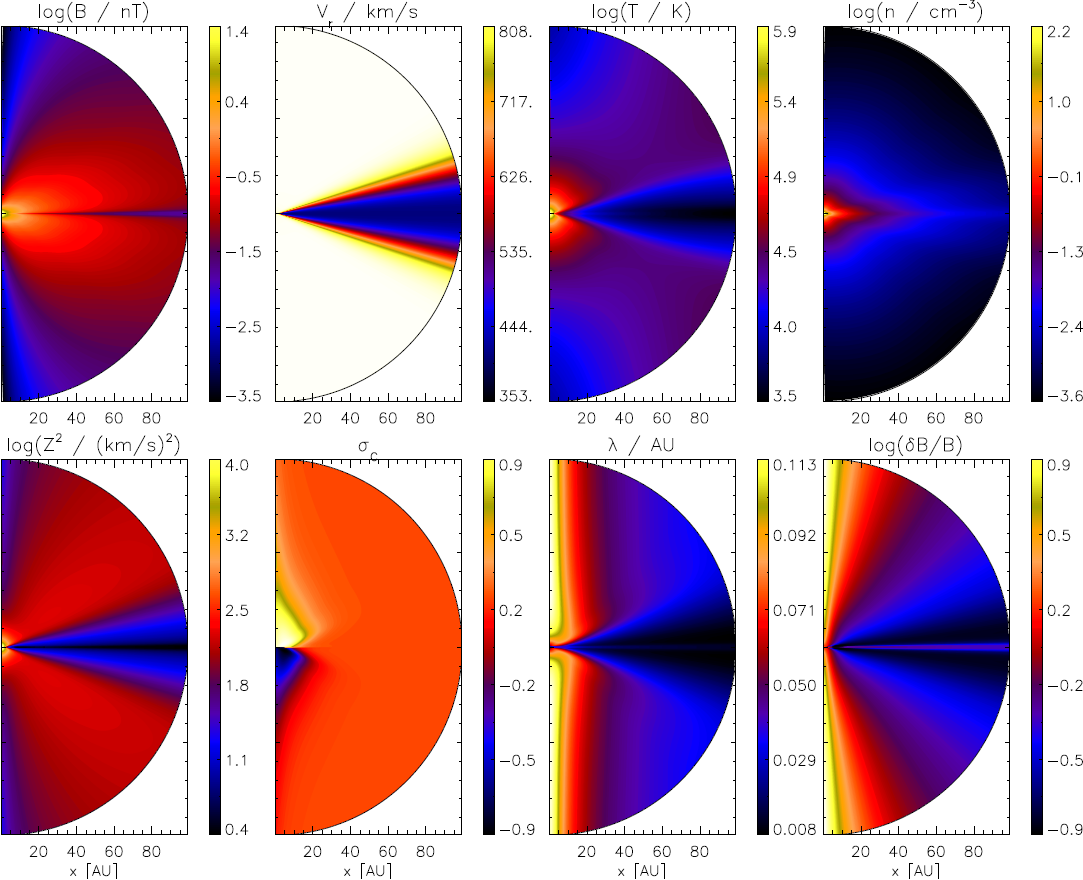}
\caption[]{Results in meridional slices for (top left to bottom right): magnetic field strength $B$, radial velocity $V_r$, temperature $T$, number density $n$, turbulent energy density $Z^2$, normalized cross-helicity $\sigma_C$, correlation length $\lambda$, and turbulence levels $\delta B/B$. }
\label{fig:Usm-slices}
\end{figure}

To validate our implementation we compare our results with those of \citet{Usmanov-etal-2011}. In order to have an equivalent set of equations the following adaptions have to be made to Equations (\ref{eq:Z2}) -- (\ref{eq:lam}): Neglecting the Alfv\'{e}n velocity and employing only the isotropization of newly born pickup ions as source for turbulence, i.e.
\begin{equation}
\langle{\bf z^\pm}\cdot{\bf S^\pm}\rangle_{pui} = \frac{\dot{E}_{pui}}{2} = \frac{1}{2}\frac{f_dUV_An_H}{n_0\tau_{ion}}\exp\left(-\frac{L_{cav}}{r}\right)~.
\end{equation}
Here, $f_D=0.25$ is the fraction of pickup ion energy transferred into excited waves, $n_H=0.1~{\rm cm}^{-3}$ is the interstellar neutral hydrogen density, $\tau_{ion}=10^6$~s is the neutral ionization time at 1~AU, $L_{cav}=8$~AU is the characteristic scale of the ionization cavity of the Sun, and $n_{sw}=5~{\rm cm}^{-3}$ is the solar wind density at 1 AU. Although neglected in the turbulence transport equations, here $V_A = B/\sqrt{\rho}$, and $U$ is the solar wind speed.\\
The background solar wind results from an untilted dipole configuration for which the boundary conditions at 0.3~AU have been fitted to be close to the ones of \citet{Usmanov-etal-2011}, as can be seen in Figure \ref{fig:Usm-init}. Thus we have a tenous and hot high-speed wind at high latitudes, while the equatorial region is occupied by a dense and cold slow-speed solar wind. The magnetic field corresponds to a Parker spiral configuration with a change of sign at the equator, resulting in a flat current-sheet there. The turbulence quantities are set accordingly to higher values in the fast wind, decreasing to smaller values in the slow wind (see also Figure \ref{fig:Usm-slices} for an overview).\\
Our boundary conditions approximate those used by Usmanov fairly well except for the transition of the magnetic field at the current sheet, which is much sharper in our case. The radial initialisation is also the same as Usmanovs.

\begin{figure}
\includegraphics[width=\textwidth]{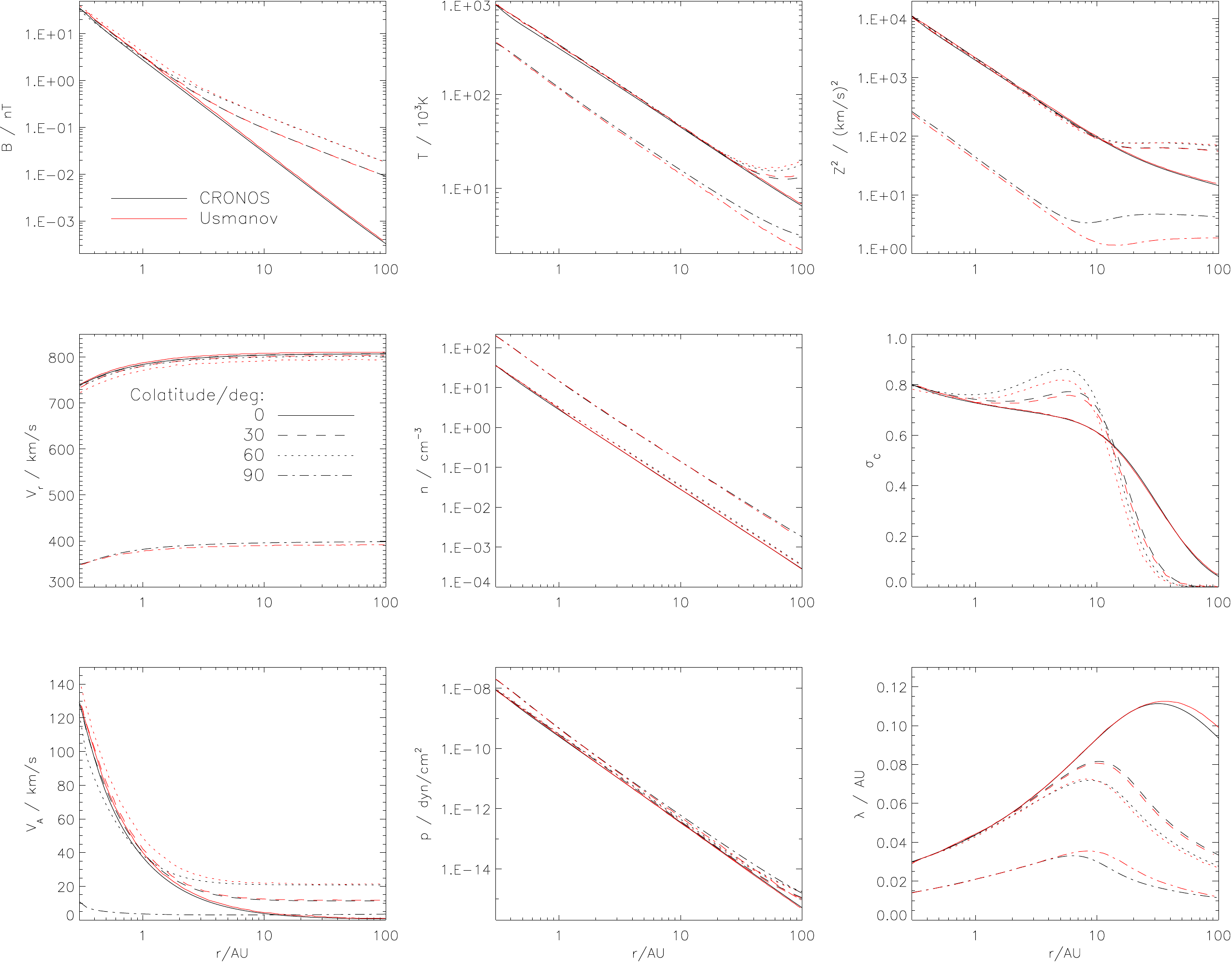}
\caption[]{Our results (black lines) in comparison with those of \citet{Usmanov-etal-2011} (red lines) for (top left to bottom right) magnetic field strength $B$, temperature $T$, turbulent energy $Z^2$, radial velocity $V_r$, number density $n$, normalized cross-helicity $\sigma_C$, Alfv\'{e}n velocity $V_A$, thermal pressure $p$, and correlation length $\lambda$.}
\label{fig:Usm-comp}
\end{figure}

The computational domain in this case covers the radial range from 0.3 AU to 100 AU with 300 cells of linearly increasing radial cell size $\Delta r\in[10,230]R_\odot$. Rotational symmetry allows us to use just one active cell (plus ghost cells) covering the azimuth $\varphi$, and the polar angle is covered with 180 cells of uniform size. The simulation is advanced in time until a steady state is reached -- which is basically the time that the slowest solar wind parcels need to reach the outer boundary. A quantitative comparison with the Usmanov results is presented in Figure \ref{fig:Usm-comp}. Shown are the radial variations of the large-scale quantities in the six left panels and the turbulence quantities in the right panels, respectively taken at colatitudes of $0^\circ$ (solid line), $30^\circ$ (dashed), $60^\circ$ (dotted) and $90^\circ$ (dashed-dotted). Note that the red curves are not those shown in \citet{Usmanov-etal-2011}, but have been obtained from A.~Usmanov (priv.~comm.) after some differences had been discovered: First, the temperature curve shown in their original paper is in disagreement with the respective curves for number density and pressure. Second, their implementation of the term ${\bf\hat{B}}\cdot({\bf\hat{B}}\cdot\nabla){\bf U}$ in spherical coordinates failed to include all geometrical source terms.\\
Our results are generally in good agreement with the newly obtained reference values, but some small deviations exist that will be addressed when briefly describing the results below: The magnetic field strength (top left) decreases as $1/r^2$ at the poles, where it is purely radial, while for lower latitudes the azimuthal component decreasing as $1/r$ becomes dominant. Slight differences with regard to the reference case are present as a result of the different boundary conditions. This is more clearly visible in the panel for the Alfv\'{e}n velocity (bottom left), however, the asymptotic values are in very good agreement. The total turbulent energy $Z^2$ (top right) decays gradually within about 10~AU, after which the turbulence generation via pickup ions becomes important and flattens the profiles. Since the pickup-ion source term is proportional to $V_A$ there is a latitude dependence resulting in a later onset of profile flattening towards high latitudes. The match between the results is excellent in the high-speed wind region, while the results at the equator beyond 10~AU are off. This is because of our constrained transport scheme involving staggered grids, where the magnetic field is stored on respective cell surfaces and the other quantities at cell centers. The current sheet is implemented best with an even number of cells in polar angle so that there is actually no cell in which $B=0$. This in turn gives no zero pickup ion term (by means of a non-vanishing $V_A$ entering), which explains the higher values for the total turbulent energy at the equator. The cell-centered quantities are actually located at half-integer values, so that in Figure \ref{fig:Usm-comp} our results are respectively offset by a half degree from those of Usmanov, which can also be expected to give some minor deviations. \\
The dissipated energy is transferred into heat so that the temperature (or pressure, related through number density (middle panels)) is higher than would be the case for an adiabatically cooling wind for given $\gamma$. This is also reflected in slightly higher terminal radial velocities (left center), where the equatorial wind is faster than in the reference case because of the dissipated higher turbulent energy as mentioned above. The cross-helicity gives the ratio between inward and outward propagating modes. As we have an inwardly directed mean magnetic field in the considered upper hemisphere, only the anti-parallel $\bf{z^+}$ modes can escape the sub-Alfv\'{e}nic region below about $20R_\odot$, so that $\sigma_C$ is close to unity. As new turbulence -- specifically also parallel propagating modes $\bf{z^-}$ -- are generated in the outer-heliosphere the ratio between the modes goes to zero. The interesting feature of increased cross-helicity between 1-10~AU is due to the additional mixing term ${\bf\hat{B}}\cdot({\bf\hat{B}}\cdot\nabla){\bf U}$. This term simplifies for a constant radial solar wind speed and Parker spiral magnetic field, resulting in ${\bf\hat{B}}\cdot({\bf\hat{B}}\cdot\nabla){\bf U}\approx(B_\varphi/B)^2(U/r)$ so that via $B_\varphi$ a latitude dependence is introduced. The effect of this term is to inhibit equipartition between forward and backward propagating modes by prefering the generation of forward propagating modes, before the onset of additional turbulence generation via pickup-ions beyond 10~AU becomes dominant. Meanwhile, the deviations with respect to the reference results for the $60^\circ$ curve are due to a lower value of magnetic field strength here as described above. The rate of dissipation is controlled by the correlation length $\lambda$ that initially rises gradually to a maximum depending on latitude and then decreases again as a result of the interplay between the right-hand terms of Equation (\ref{eq:lam}). The breaks at about 10~AU are again caused by the pickup-ion term, indicating that sources of turbulence tend to decrease the correlation length, which will also be addressed when including shear driving below. \\
The overview presented in Figure \ref{fig:Usm-slices} also shows the ratio of turbulent magnetic field to mean magnetic field $\delta B/B$ as an additional quantity, which can be calculated via
\begin{equation}
	\delta B^2 = \langle b^2 \rangle  = \frac{\rho Z^2}{2}(1-\sigma_D)
\end{equation}
for given (constant) $\sigma_D$. This quantity bears great importance for the transport of energetic particles in the heliosphere as it is a measure for the efficiency of diffusion \citep[see, e.g.,][]{Manuel-etal-2014}. The results show that although the turbulent energy is low in polar regions, the magnetic field strength there diminishes more rapidly and results in high diffusion levels that gradually decrease towards the ecliptic.

\section{Model extension}
\label{sec:extension}
It is stated in \citet{Usmanov-etal-2011} that their model does not seem to require additional source terms to account for the effects of turbulence driven by shear. However, as acknowledged in \citet{Usmanov-etal-2014}, this is not the case. Furthermore, this is also pointed out in Appendix A of \citet{Zank-etal-2012a}, where it is noted that these kind of turbulence transport models do not capture turbulence driven by shear terms due to the imposed structural similarity closure relations, such that instead these terms have to be explicitly accounted for via additional source terms. \citet{Usmanov-etal-2014} incorporated shear effects with an eddy viscosity approximation, which we intend to adopt in our implementation for future studies as well. For now we include the required terms in a similar fashion as in \citet{Zank-etal-1996} and \citet{Breech-etal-2008}, but in improvement to these models we compute the gradients in the solar wind speed self-consistently from the background field via
\begin{equation}
\langle{\bf z^\pm}\cdot{\bf S^\pm}\rangle_{sh} = \frac{1}{2}Z^2\underbrace{C_{sh}\left|\frac{1}{r}\left(\partial_\vartheta+\frac{1}{\sin(\vartheta)}\partial_\varphi\right)|{\bf U}|\right|}_{\hat{C}_{sh}}~.
\label{eq:shear}
\end{equation}
We use a value of $C_{sh}=0.5$, which is commonly taken at high latitudes (no shear) to match observational results \citep{Breech-etal-2005}. Previous studies (e.g.~\citet{Breech-etal-2008,Engelbrecht-Burger-2013}) have estimated typical values for gradients in the solar wind speed and absorbed them in the definition for $\hat{C}_{sh}$ so that it is varying with position. Here, we treat $C_{sh}$ as a constant and get an equivalent expression for $\hat{C}_{sh}$ as indicated in Equation (\ref{eq:shear}).\\
\begin{figure}
\includegraphics[width=\textwidth]{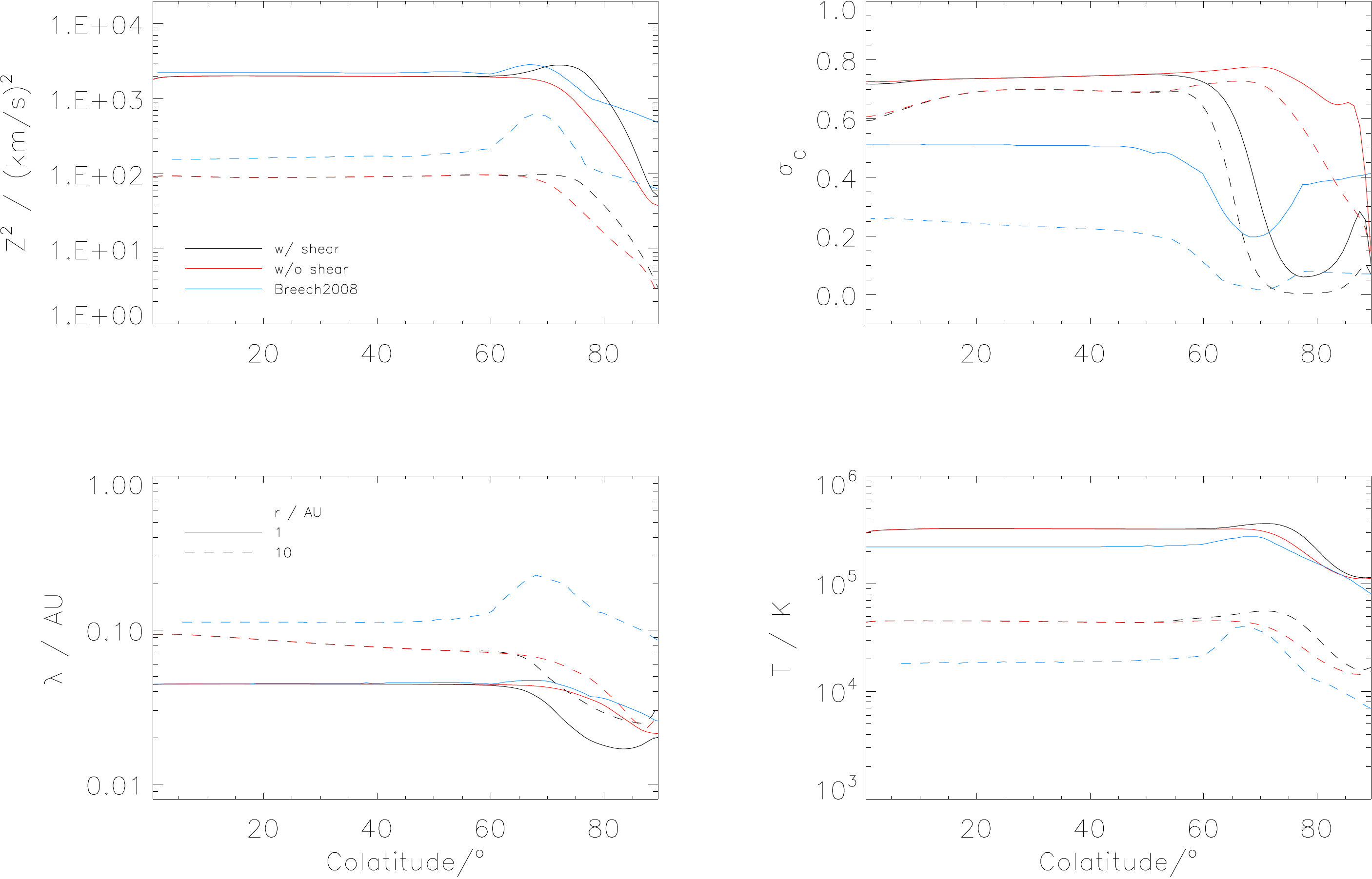}
\caption[]{Turbulence quantities and temperature in meridional slices at different heliocentric distances (solid: 1AU, dashed: 10AU). The black (red) lines show the results (not) including the shear term, while the blue lines are taken from \citet{Breech-etal-2008}. }
\label{fig:Shear-comp}
\end{figure}
Figure \ref{fig:Shear-comp} shows a comparison of simulations using the above setup with and without the shear term in meridional slices at different heliocentric distances (solid: 1AU, dashed: 10AU). The black (red) lines show the results (not) including the shear term, while the blue lines are taken from \citet{Breech-etal-2008}. As expected, differences arise mainly in the transition region between slow and fast solar wind (at about 70 -- $80^\circ$), which gives rise to shear. Considering the 1AU slices (solid lines) first, it can be seen that the shear term leads to (otherwise absent) enhancements in turbulent energy, which are comparable to the enhancements in the Breech results, where the latter are at slightly lower colatitudes due to different modeling of the transition region. Meanwhile, the cross-helicity is now strongly decreased in this region due to the newly generated turbulence, which is also qualitatively visible in the Breech data, who however used lower boundary values for the cross-helicity so that it is initially lower. While the correlation length is enhanced in the Breech model, it is decreasing in the shear region in our model. This is due to our inclusion of the term in the respective Equation (\ref{eq:lam}) for the correlation length, where as mentioned above with regard to the pickup-ion source term, the generation of turbulence leads to decreased correlation lengths. The shear term is absent in the evolution equation for $\lambda$ in the Breech model, where it is assumed that shear drives turbulence at all scales. As our starting point for the turbulence transport equations is the model by \citet{Zank-etal-2012a}, who include this term also in the correlation length equation, we will maintain it as well. While at 1AU the resulting temperature enhancement is similar to the one in Breech (different boundary values again cause an overall lower temperature here), the results at larger radial distances show some differences: in our model the turbulent energy just barely peaks at the transition region and the temperature is only slightly enhanced as well, whereas in the Breech model the peak becomes ever more pronounced. This is the result of an interplay between the generation of turbulence due to shear, which weakens with radial distance, and its dissipation controlled by the correlation length. A more detailed study addressing the effect of shear on the correlation length might be in order to clarify proper modeling. We want to mention though that Ulysses data for solar minimum conditions \citep[plate 6 in][]{McComas-etal-2000} do not seem to show evidence for a strongly enhanced temperature in this region.    
  
\begin{figure}
\includegraphics[width=\textwidth]{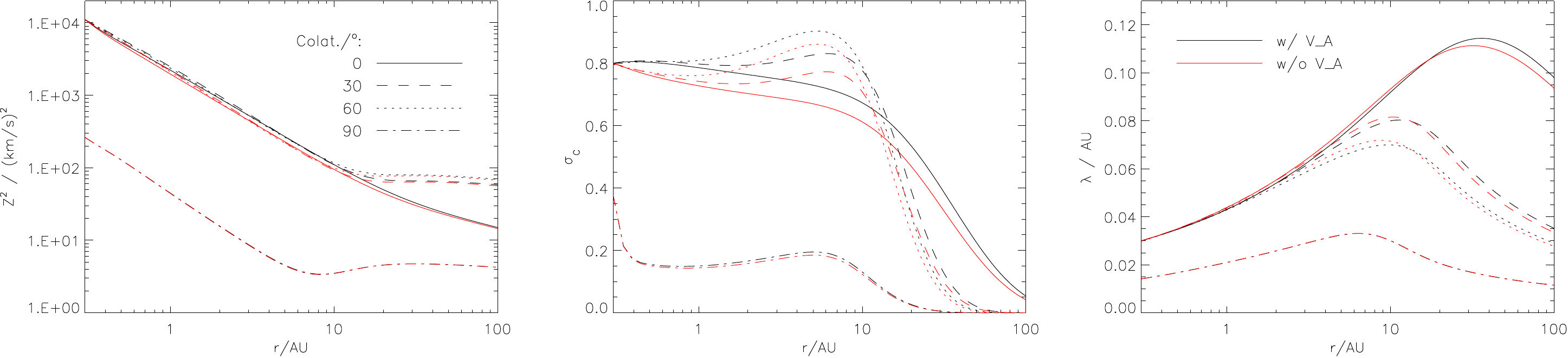}
\caption[]{Results for the turbulence quantities with (black lines) and without (red lines) including the Alfv\'{e}n velocity terms. }
\label{fig:VA-comp}
\end{figure}
The model can be readily extended to include terms involving the Alfv\'{e}n velocity. We maintain the same boundary conditions as before and the results are compared to the ones neglecting Alfv\'{e}n velocity in Figure \ref{fig:VA-comp}. The effect on the large-scale quantities is negligible and, thus, not shown. The turbulence quantities also show only minor differences: The turbulent energy tends to be up to about 1.5 times higher at high latitudes and the correlation length is almost unaffected. Near the equator the results are almost identical for all quantities. The normalized cross-helicity also tends to stay about 10\% larger when including the additional terms. These deviations occur close to the inner boundary, where the Alfv\'{e}n speed is highest, and the resulting higher cross-helicity values are convected to larger radial distances. This effect is expected to be more prominent when moving the inner boundary closer to the Sun, as will be done in the subsequent section. For an inner boundary at about 0.3~AU as considered so far, the neglect of the Alfv\'{e}n velocity terms is therefore justifiable to some extent, but the effect of enhanced cross-helicity could be incorporated in models neglecting these terms by increasing the boundary values accordingly.

\section{Application to the inner heliosphere and transient structures}
\label{sec:innerheliosphere}
The turbulence transport equations (\ref{eq:Z2}) -- (\ref{eq:lam}) now also accomodate the effect of shear driving by means of Equation (\ref{eq:shear}) so that the model can be applied to solar wind transients with arbitrary gradients in solar wind speed. In Section \ref{sec:CME} we apply the above set of equations to a "toy-model" CME and study the results in comparison to the model not including turbulence. Furthermore, including terms involving the Alfv\'{e}n velocity enables us to move the inner boundary closer to the Sun. In the following the inner boundary is located at 0.1~AU, which is usually just beyond the Alfv\'{e}n critical radius. This is a suitable choice for the following reasons: (i) The turbulence transport model is only partly appropriate for sub-Alfv\'{e}nic regions such as the corona or the heliosheath. Specifically, the dissipation terms are subtle to model for low-beta regions and still need to be properly adapted for such cases (G.~Zank, priv.~comm.; see also \citet{Zank-etal-2012a}). (ii) Coronal MHD models are much more complex and are computationally quite expensive. A simplified model of the corona is the WSA model that is frequently used to derive boundary conditions for heliospheric MHD simulations such as in our previous work \citep{Wiengarten-etal-2014} or in the ENLIL code \citep{Odstrcil-etal-2004}, which is in operation at the space weather prediction center. Here, the interface between WSA and MHD is usually located at 0.1~AU, so that especially for future purposes, where we will aim to perform the simulations with input from the WSA model, this would be the obvious choice. (iii) When applying our model to interplanetary disturbances such as CMEs, it is desirable to catch as much of their evolution self-consistently, which demands to put the inner boundary as close to the Sun as possible. Given the above caveats, this currently cannot be done in a more self-consistent manner by triggering a reconnection event at the coronal base (as in, e.g., \citet{Manchester-etal-2005, Kozarev-etal-2013}), but instead an estimate for CME properties at intermediate distances has to be found (see Section \ref{sec:CME}).\\
\\
In what follows the simulation domain is restricted to the radial extent $r\in[0.1,1.2]$~AU and we do not cover the polar coordinate singularities to avoid small time steps. Therefore, $\vartheta\in[0.1,0.9]\pi$. Furthermore, to save computing time the azimuthal extent is restricted to $\varphi\in[0.5,1.5]\pi$ as the center of the CME will be located at $[\vartheta,\varphi]_{CME}=[0.5,1]\pi$, and during its subsequent evolution it does not reach the outer azimuthal boundaries in this setup. The applied resolution is $[\Delta r,\Delta\vartheta,\Delta\varphi]=[0.5R_\odot,1^\circ,1^\circ]$ corresponding to gridcells per direction as $[N_r,N_\vartheta,N_\varphi]=[480,144,180]$. 

\subsection{Quiet Solar Wind}
\begin{figure}
\includegraphics[width=\textwidth]{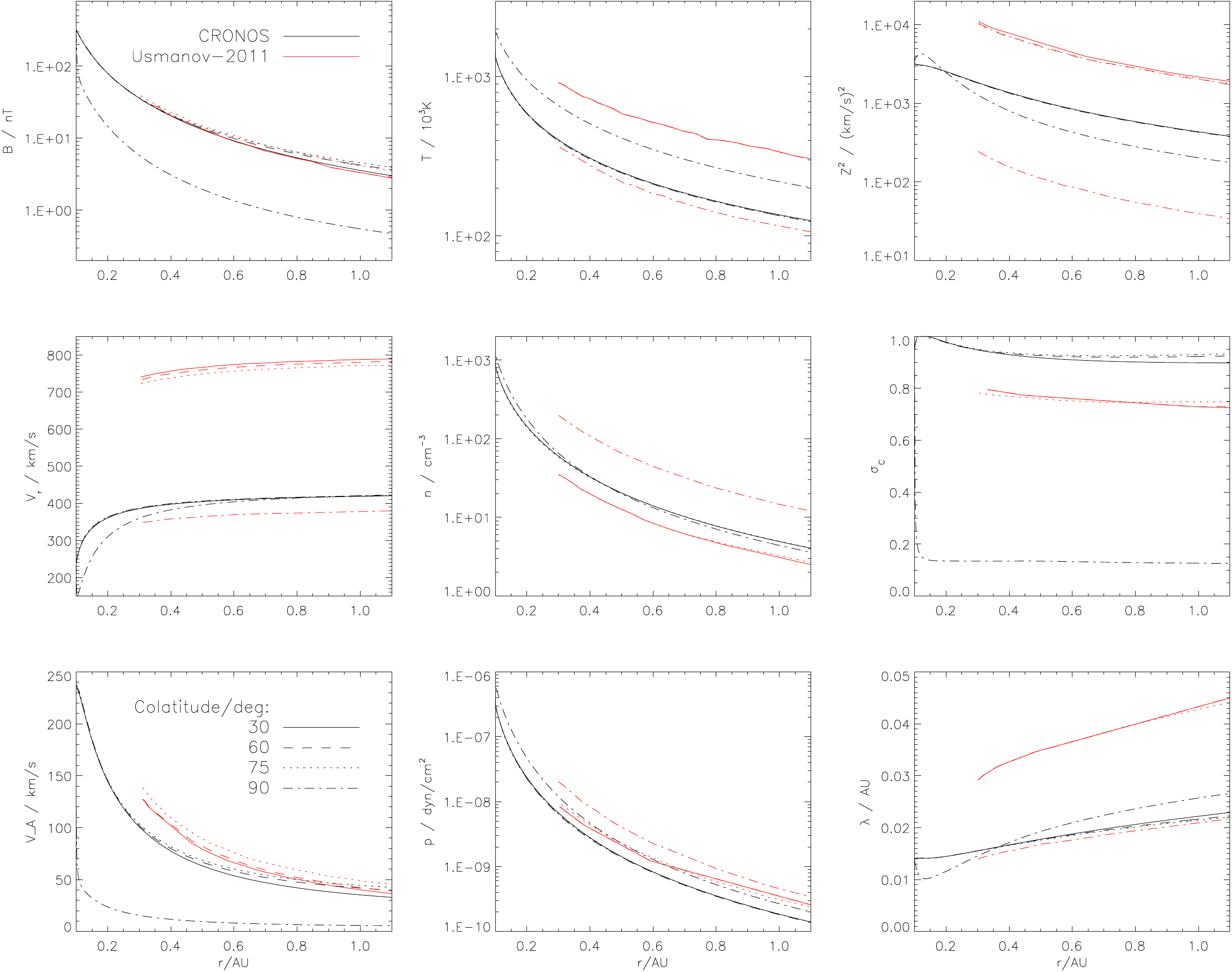}
\caption[]{Results for the quiet solar wind at different colatitudes in the same format as Figure \ref{fig:Usm-comp}.  }
\label{fig:quiet-turb}
\end{figure}

We first describe the quiet background solar wind, into which a CME will be injected. CMEs occur more frequently during periods of solar maximum, which is characterized by a highly tilted current sheet and disapperance of the region of fast polar solar wind, so that instead a slow wind is present at all latitudes. For simplicity, and since we leave out the polar coordinate singularity, we resort to a simple flat current sheet. This also allows us to keep the analysis of the results relatively tractable.\\
We estimated respective boundary conditions at $0.1$AU for the large-scale quantities from typical values used in our previous work employing the WSA model \citep{Wiengarten-etal-2014}. The boundary values and the resulting radial evolution at selected colatitudes are shown in Figure \ref{fig:quiet-turb} (black lines), where for orientation the findings of \citet{Usmanov-etal-2011} (red lines) from 0.3 AU onwards are shown as well. The resulting configuration is similar to the equatorial slow speed region from the previous section but with slightly enhanced speeds, higher temperature, and lower density, while the magnetic field is the same as before. For the turbulent energy and the correlation length we use boundary values that also give similar radial profiles as shown previously, but for the cross-helicity it should be assumed that just beyond the Alfv\'{e}nic critical radius there are almost only forward propagating modes, so that $|\sigma_C|$ should be close to unity and we choose $|\sigma_C|(0.1AU)=0.95$. As shown above, including the Alfv\'{e}n velocity terms in the model tends to retain cross-helicity values closer to unity, and since there are no additional sources of turbulence so far the cross-helicity now remains larger at and beyond 0.3 AU in contrast to the Usmanov values. 
   

\subsection{Pertubation by a CME}
\label{sec:CME}
\subsubsection{Initialization}
As the computational domain does not extend down to the corona, where CMEs are thought to be triggered via reconnection events, we do not use an injection scheme that inserts out-of-equilibrium flux ropes as done in coronal models \citep[e.g.,][]{Manchester-etal-2005, Kozarev-etal-2013}, but we estimate typical CME properties at the intermediate distance of $r_0 = 0.1$~AU from simulations performed by \citet{Kleimann-etal-2009}. We find that the shape of the CME in these simulations remains similar to the initialized one, as was also reported in previous studies and motivated the Cone model \citep{Zhao-etal-2002}, which is also used to initiate CMEs in the ENLIL setup \citep{Odstrcil-etal-2004}. The almost radial propagation of CMEs through the corona allows for an easy geometrical estimation of its properties a few solar radii away from the Sun. While for space weather forecasting studies such estimates are based on observations, here we consider a "toy model" with idealized properties. We first define the normalized angular distance of a given point $[\vartheta,\varphi]$ on the spherical inner boundary to the center of the CME onset site $[\vartheta,\varphi]_{cme}=[0.5,1]\pi$ as 
\begin{equation}
a = \arccos [X_{0,cme}\sin(\vartheta)\cos(\varphi)+Y_{0,cme}\sin(\vartheta)\sin(\varphi)+Z_{0,cme}\cos(\vartheta)]/\delta_{cme} ~,
\end{equation}
where $[X,Y,Z]_{0,cme} = [\sin(\vartheta_{cme})\cos(\varphi_{cme}), \sin(\vartheta_{cme})\sin(\varphi_{cme}),\cos(\vartheta_{cme})]$ are the Cartesian coordinates of the site's center, and $\delta_{cme} = \pi/8$ denotes its angular radius. Within this circular patch characterized by $a(\vartheta,\varphi)\leq1$, we raise the large-scale fluid quantities during the CME initialisation according to $[v_r, T, n]_{cme} = [8v_r, 4T, 1.5n]_{quiet}\cos(0.5\pi\cdot a)f(t)$, so that the enhancement peaks at the center and decreases towards the edges of the circular area. For the time dependence we choose a linearly decreasing function $f(t)$ from maximal values at $t_{cme}$ back to quiet values at $t_{cme}+\delta t_{cme}$ with a duration of $\delta t_{cme} = 5t_0$, where our normalization value for time $t_0=3191$s. Furthermore, the onset time $t_{cme}=200t_0$ is chosen as such that the initial quiet conditions have reached steady-state.\\
Although the magnetic field strength could be simply raised as the other quantities, this would not change the fields topology (Parker spiral), which we assume to be affected qualitatively by the CME's onset. We therefore prescribe an additional strong $B_\vartheta$ component as to get field-lines wrapping around the central region of the CME. To preserve the solenoidality constraint we directly prescribe the vector potential
\begin{equation}
	{\bf A} = \left(\begin{array}{c} A_r \\ A_\vartheta \\ A_\varphi \end{array}\right) = \left(\begin{array}{c} 0 \\ -B_0r_0^2\sin(\vartheta)(\varphi/r+\Omega/V_r) \\ -B_{0,cme}r_0\cos(0.5\pi\cdot a)f(t) \end{array}\right)~,
\end{equation}
where $A_\vartheta$ gives rise to the Parker spiral magnetic field, while $A_\varphi$ results in the desired field lines wrapping around the CME (see Figure \ref{fig:cme-para}). \\
Meanwhile, we leave the turbulence quantities at the quiet solar wind boundary values. This is probably not a realistic assumption as the evolution in the sub-Alfv\'{e}nic region should have an impact on the turbulence quantities as well. However, on the one hand there is little to no observational data of turbulence associated with CMEs this close to the Sun, and on the other hand, prescribing respective estimates as boundary conditions for the turbulence different from the quiet wind ones would make it more difficult to analyse the self-consistent evolution of these quantities beyond 0.1~AU, i.e.~the influence of different boundary conditions and the evolution due to the governing equations would be impossible to disentangle. \\

\subsubsection{Results and impact on turbulence quantities}
\begin{figure}
\includegraphics[width=\textwidth]{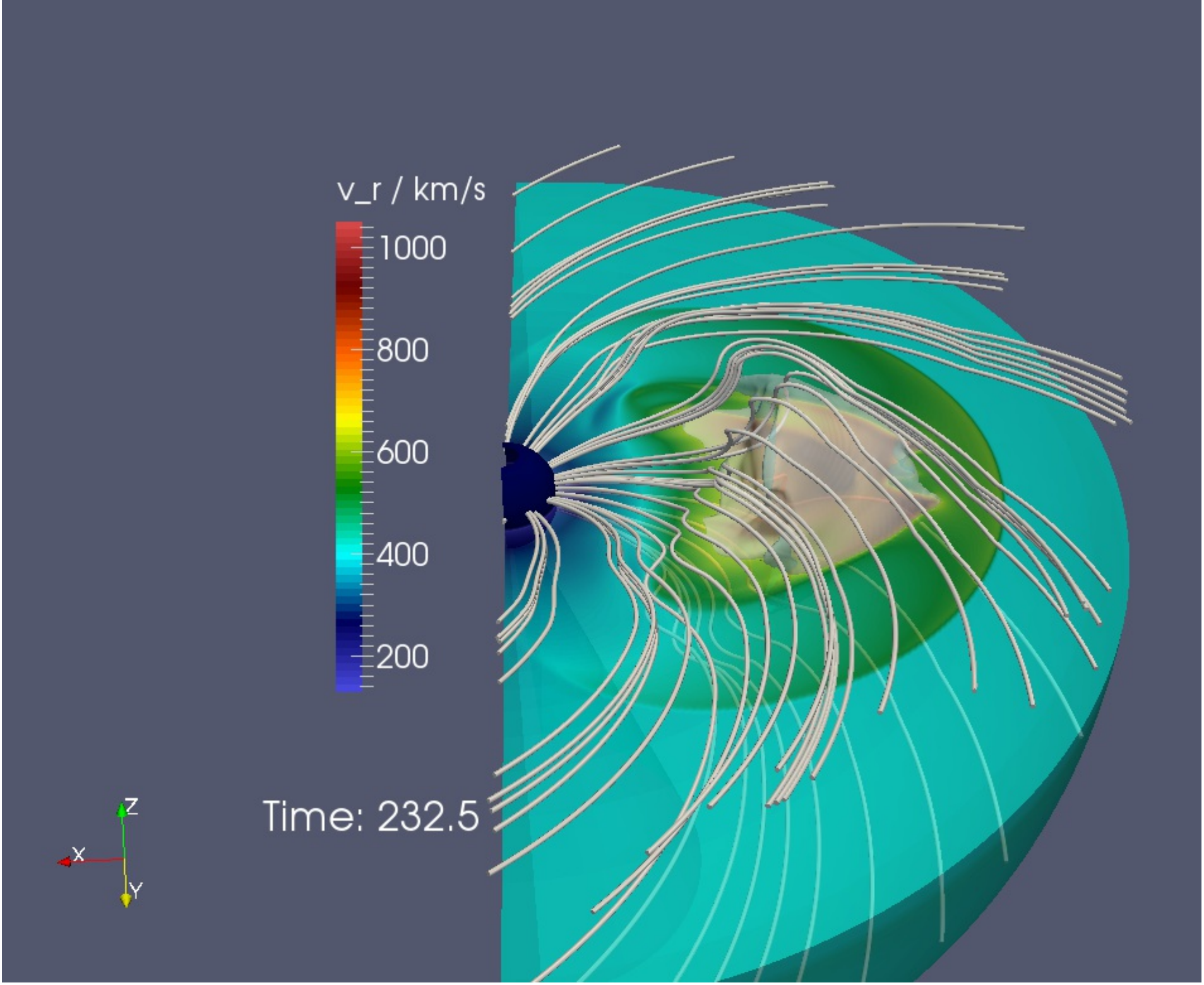}
\caption[]{3D visualization of CME with magnetic field lines. The center sphere is the inner boundary, while the computational domain is clipped at the equator and made opaque. The color coding is for radial speed and shows the sheath and core region of the CME (see text), while the whitish opaque shape is the contour of $V_r=650$km/s that shows the 3D extent of the core region of the CME. The time evolution as an animation of this figure is available in the supplementary material.}
\label{fig:cme-para}
\end{figure}

\begin{figure}
\includegraphics[width=\textwidth]{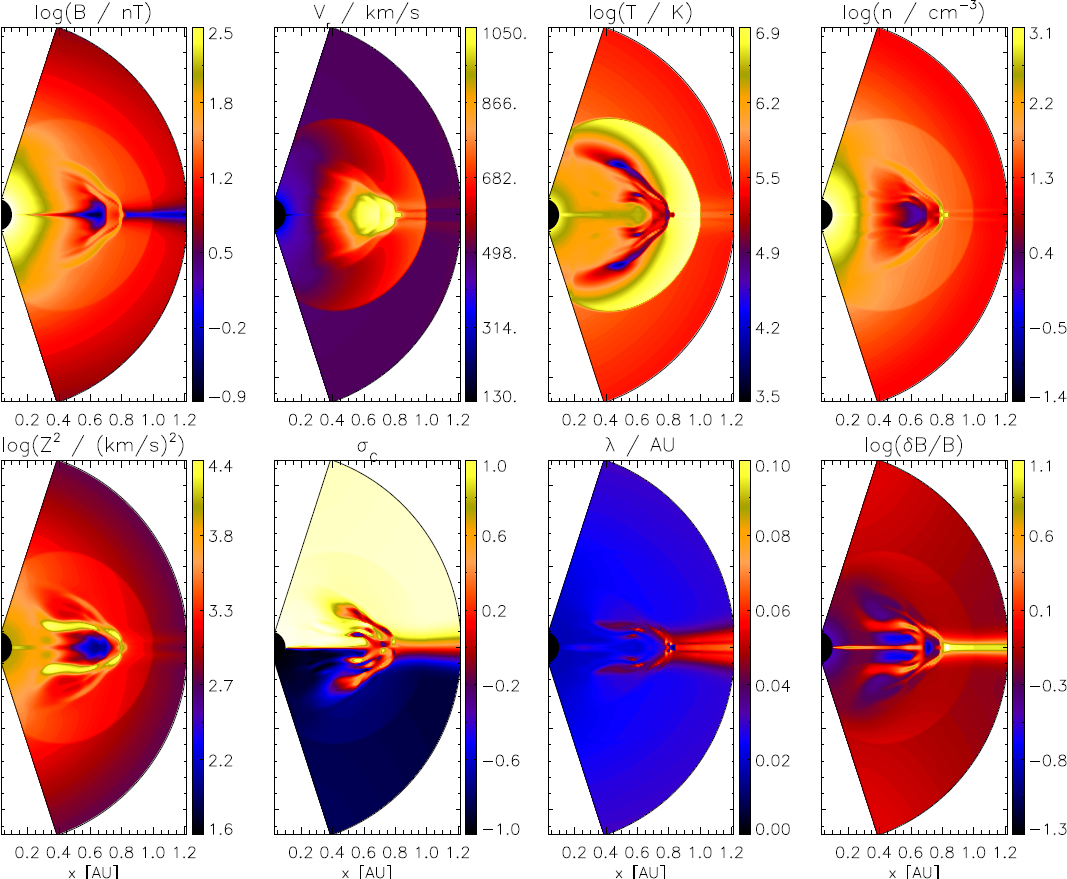}
\caption[]{Results including turbulence at $t=32.5t_0$ after CME injection in meridional slices at $\varphi=\pi$. The time evolution as an animation of this figure is available in the supplementary material. }
\label{fig:cme-turb}
\end{figure}

\begin{figure}
\includegraphics[width=\textwidth]{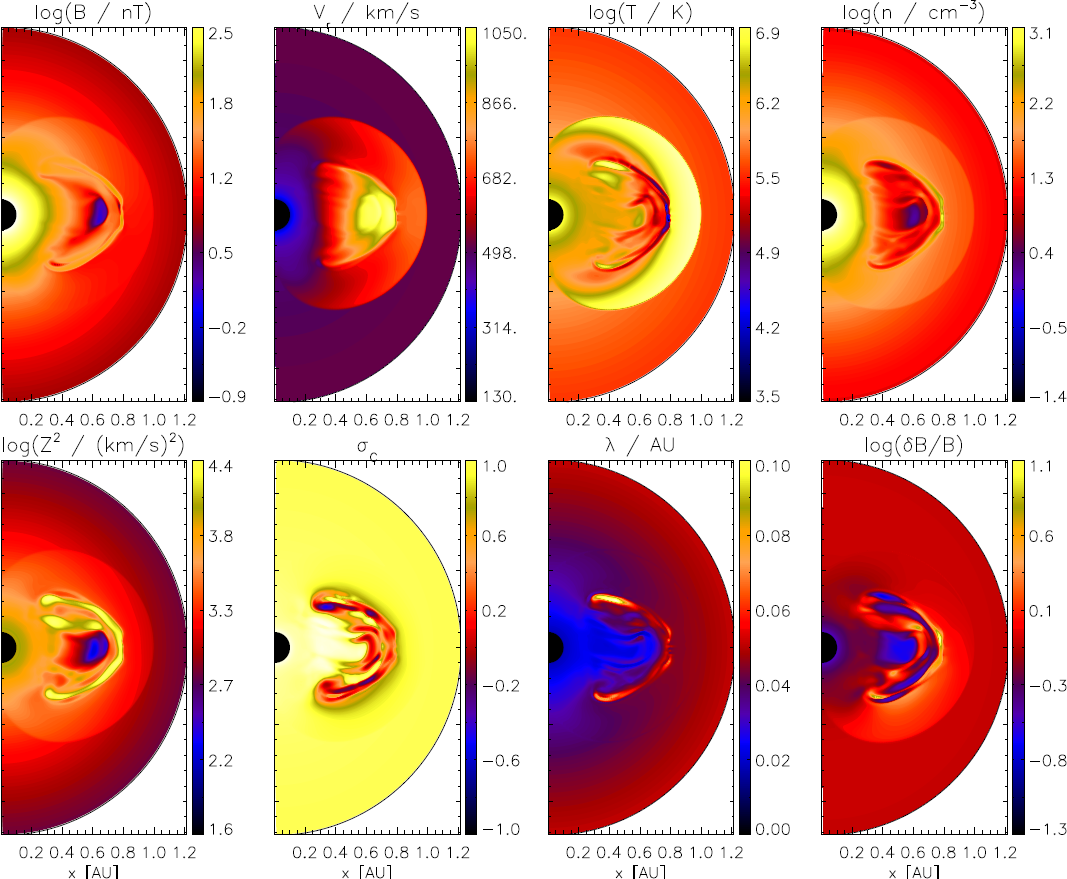}
\caption[]{Results including turbulence at $t=32.5t_0$ after CME injection in azimuthal slices at $\vartheta=85^\circ$. The time evolution as an animation of this figure is available in the supplementary material. }
\label{fig:cme-turb-azim}
\end{figure}

An overview of the resulting 3D structure is shown in Figure \ref{fig:cme-para}. The field lines show the typical Parker spiral pattern far away from the CME, while close by they wrap around it. The CME can be partitioned into a compact central core region containing the high speed ejecta and a surrounding sheath region bounded by a shock driven by the CME.
Due to the wealth in structure involved, we show results in 2D slices. First, we present meridional slices at $\varphi=\pi$ in Figure \ref{fig:cme-turb} so that they conincide with the azimuthal symmetry axis of the {\it initial} CME. The evolution in time is available as a movie in the supplementary material. Here, we show an adapted snapshot of the movie at $t_{ss}=32.5t_0$ after initialization, which is the time at which the sheath region reaches 1~AU. The core and sheath region are most clearly seen in the large scale quantities (top row), but in particular in the radial velocity data, whereas in most of the quantities more complex structures are visible as well. In general, structural variations occur mainly close to or within the elongated core region, while the almost circular sheath region is rather homogeneous in comparison. The sheath region is characterized by modest enhancements compared to the quiet conditions (visible beyond $\approx$ 1~AU) in density and magnetic field strength, while strong enhancements of $v_{r,sheath}\approx700$~km/s and $T_{sheath}\approx10^7$~K are found. The turbulent energy is also increased in this region, and also more so than the magnetic field strength, which follows from the visible enhancements in $\delta B/B$. The cross-helicity is slightly reduced, while the correlation length seems to exhibit almost no changes within the sheath region. \\
The core region partly overlaps with the current sheet affected equator. The current sheet is not ideally resolved, which would require adaptive mesh refinement, which, however, is not yet available in the {\sc Cronos} framework. Therefore, close to the current sheet we get strips of increased or depleted values that are overestimated in this model, but this does not affect the results offset from the equator too much. \\
While the sheath region is clearly discernible and has the same extent in all quantities, the core region is not similar in all quantities. The patch containing the highest velocity values reflects different behavior in the other quantities, where basically two regions can be distinguished: On the one hand there is a compression at the leading edge, also towards higher latitudes, with respective elevations in magnetic field strength, density, and turbulent energy, while on the other hand a rarefaction region trails the core, where depleted values are present. Besides these compressional effects there is clearly a distinct band of high turbulent energy due to shear around the edges of the high-speed central region. Respective structures are also visible in the panels for the correlation length and for the cross-helicity, where the latter is generally speaking closer to zero in most parts affected by the CME but also shows some additional structure, probably due to the complex magnetic field geometry. Finally, the panel for $\delta B/B$ reveals enhanced values in the sheath and around the core region, while the rarefaction region exhibits reduced turbulence levels. The respective effects on the perpendicular and parallel mean free path of energetic particles is twofold, as the former is proportional to $\delta B/B$, while the latter is antiproportional to it. This can lead to counterintuitive results, where the presence of enhanced turbulence can even facilitate the transport of energetic particles as found by \citet{Guo-Florinski-2014} for CIRs. \\
While the meridional slices are symmetric with respect to the equatorial current sheet, this cannot be expected for azimuthal slices because of the symmetry breaking in the spiral structure of the magnetic field. Figure \ref{fig:cme-turb-azim} shows respective results in azimuthal slices at $\vartheta=85^\circ$, where this colatitude is chosen in order to show the behavior not directly at the current sheet, whose influence is overestimated in this model. These slices are oriented in accordance with the orientation in Figure \ref{fig:cme-para}. From the field lines shown there, it is clear that the magnetic field topology is different at the upper edge of the CME, where the field lines are compressed in a sense according to the initial bending direction of the Parker spiral, as compared to the lower edge where field lines are compressed in a sense opposite to the spiral bend. While many features remain similar to the ones described above for the meridional perspective, there are some additional features due to the symmetry breaking: The core region's tail is bent towards the lower edge, while the sheath region is rather unaffected, i.e.~remains quite symmetric in the hydrodynamic quantites ($V_r$, $T$, $n$), but as discussed above the magnetic field compression in the sheath region is stronger at the upper edge, which also affects the turbulence quantities: At the upper edge of the core region there is an additional feature of reduced cross-helicity and enhanced turbulence, which is also reflected by a respective feature in temperature, whereas these structures are absent at the lower edge. A striking azimuthal structure for the diffusion levels is found, as the upper edge of the CME including the sheath region there show relatively small values, while towards the nose and the lower edge the diffusion levels are markedly higher. This represents an intriguing feature for a subsequent study of energetic particle propagation.\\
To the best of our knowledge the observations of turbulence in or associated with CMEs
are rather limited and use data obtained at 1~AU \citep[e.g.][]{Ruzmaikin-etal-1997}. The
first study that derived quantitative estimates of magnetic turbulence levels near CME
fronts was presented by \citet{Subramanian-etal-2009}. Interestingly our findings are 
within the limits provided by these estimates.\\


\subsubsection{Impact of turbulence quantities on CME}
\begin{figure}
\includegraphics[width=\textwidth]{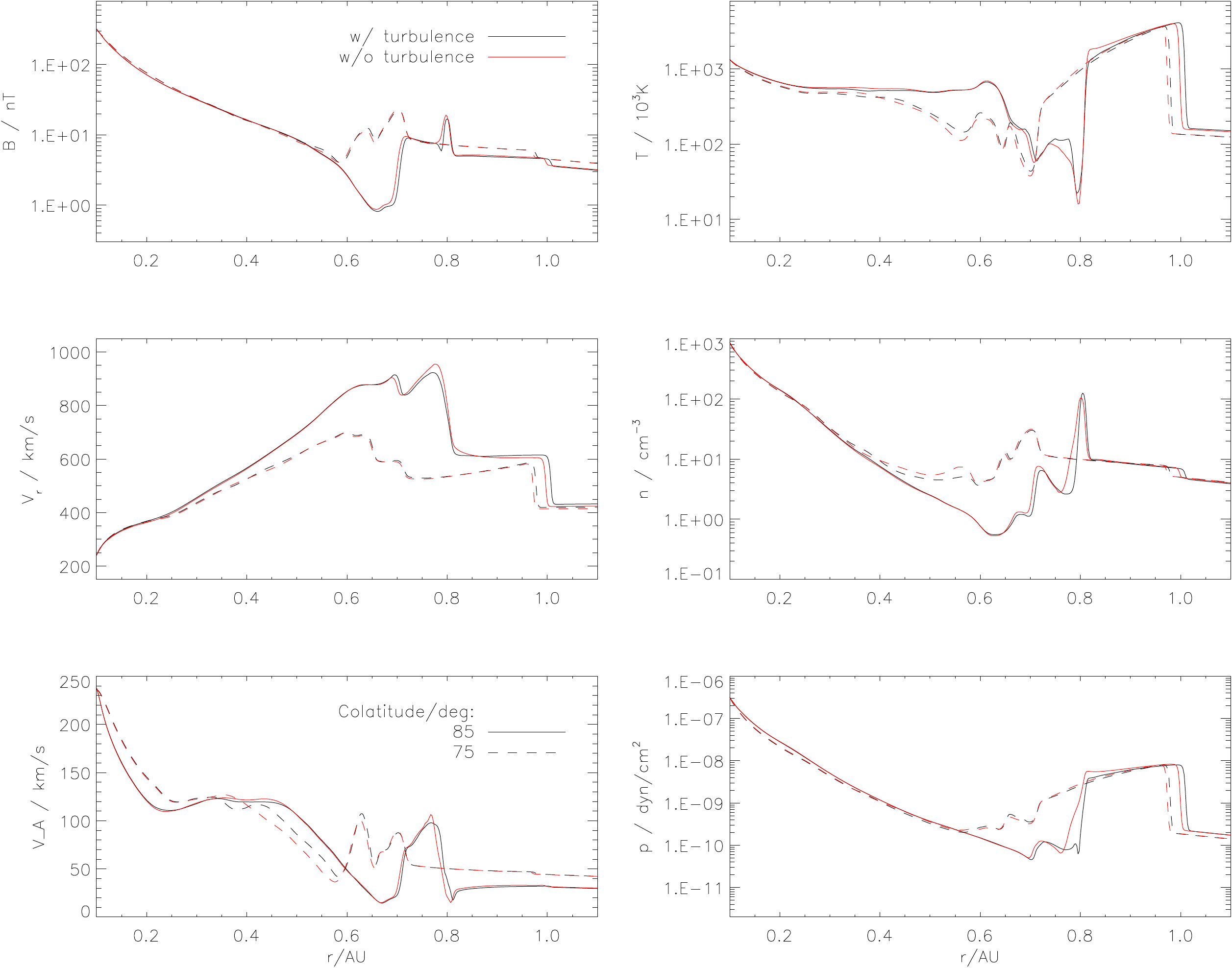}
\caption[]{Comparison of large-scale quantities for simulations with (black) and without turbulence (red). }
\label{fig:cme-comp}
\end{figure}

In the previous section we described the impact of disturbances in the large-scale flow via a CME on the turbulence quantities. We now study the effect of the turbulence quantities on the disturbed large scale flow by comparing the simulations discussed above with simulations where the turbulence is switched off, but the remaining setup is maintained. \\
We omit to show 2D slices as above since the results remain remarkably similar. Instead, Figure \ref{fig:cme-comp} shows radial profiles for the large-scale quantities at $\varphi=\pi$ and colatitudes of $85^\circ$ (solid lines) and $75^\circ$ (dashed lines) for the case with (black lines) and without turbulence (red lines). The deviations are very small in almost all regions, but a notable difference is the extent of the sheath region, which is marked, e.g., by the largest temperature values. The additional heating due to enhanced turbulence there results in a slightly more extended sheath region, but the effect can be considered small enough so that it is negligible for general CME propagation studies.


\newpage
\section{Summary and Outlook}
\label{sec:conclusions}
In this paper we presented our implementation of turbulence transport coupled to the Reynolds-averaged ideal MHD equations in the framework of the {\sc Cronos} code. We followed the work of \citet{Usmanov-etal-2011} and validated our findings by comparing results with these authors' findings. Their original model was extended to be applicable to regions of solar wind speeds that do not have to greatly exceed the Alfv\'{e}n speed, which we achieved by simplifying the more general turbulence transport equations of \citet{Zank-etal-2012a}. It was shown that beyond radial distances of $0.3$ AU the neglect of such terms is usually justified, but including them results in a slower decrease of normalized cross-helicitiy values, i.e.~the generation of backward propagating modes is somewhat inhibited. This effect is stronger when moving the inner boundary of the simulations closer to the Sun, where the Alfv\'{e}n speed is no longer small compared to the solar wind speed. Some additional terms that were inappropriately absent in the \citet{Usmanov-etal-2011} model have also been included in the present work: (i) the effect of turbulence driven by shear was introduced via respective ad-hoc terms, whose strength was estimated from comparisons with the work of \citet{Breech-etal-2008}. Such an ad-hoc approach has to  be taken because the structural similarity assumption, commonly made to achieve closure in the turbulence transport equations, prevents the self-consistent formulation of shear driving. Recently, \citet{Usmanov-etal-2014} removed this constraint by employing an eddy-viscosity approximation, which we plan to include in the future as well. (ii) The mixing term ${\bf\hat{B}}\cdot({\bf\hat{B}}\cdot\nabla){\bf U}$ was now correctly implemented that results in a (latitude-dependent) effect on the normalized cross-helicity, which is now even increasing towards about 10~AU, after which turbulence driven by ionization of pick-up ions becomes dominant and results in equipartition between forward and backward propagating modes.\\
While most previous studies of turbulence in the solar wind use prescribed solar wind conditions such as a constant wind speed and a simple Parker spiral magnetic field, our approach allows for the self-consistent evolution of turbulence in a time-dependent and disturbed solar wind. Together with our extension towards non-highly-super-Alfv\'{e}nic solar wind speeds, we are therefore able to move the inner boundary of our simulations closer to the Sun and consider the effects of transient structures on the turbulence quantities. While the effect of CIRs was already addressed in previous studies \citep{Usmanov-etal-2012,Usmanov-etal-2014}, we concentrate on CMEs, here, by using an injection scheme similar to the Cone model \citep{Zhao-etal-2002}, estimating typical CME properties at intermediate solar distances. Instead of trying to recreate a specific observed CME, we consider a relatively easy toy-model kind of scenario to study general effects and reserve a more detailed event study with respective comparison to spacecraft data for future work. While the large-scale quantities show typical structures such as a core and sheath region of the CME as found in previous studies \citep[e.g.~][]{Manchester-etal-2005, Kozarev-etal-2013}, here for the first time, we show the respective 3D structure of turbulence quantities of CMEs. These show enhanced turbulence levels and reduced cross-helicity values caused by the CME. Azimuthal symmetry is broken by the disturbed magnetic field and results in azimuthally varying diffusion levels, which will provide an interesting feature to study with energetic particle propagation models.\\
We also investigated possible back reactions of turbulence on the large-scale flow by comparing with identical simulations save for the inclusion of turbulence. We find a slightly larger CME sheath region due to additional heating via the turbulent cascade, but in general the effect on the propagation of the CME is negligible and, therefore, does not have to be incorporated in models interested in the large-scale flow only. \\
Besides the future directions mentioned above, we will in a forthcoming study especially focus on implementing the full set of turbulence transport equations of \citet{Zank-etal-2012a}, which allow for a variable energy difference $\sigma_D$ (thus far assumed to be constant) and different correlation lengths for the forward and backward propagating modes as well as for the energy difference, whereas so far a single correlation length was adopted. We also plan to explore the possibility to include two component turbulence \citep{Oughton-etal-2011}. Furthermore, the background solar wind can be modeled more realistically with observationally based boundary conditions derived with the WSA model as reported in our previous work \citep{Wiengarten-etal-2014}, with which we will be able to model inner-heliospheric conditions for the background solar wind and associated turbulence. Both can then be directly compared to spacecraft observations and subsequently will serve as input to energetic particle propagation models.

\acknowledgments


Many thanks are in order to A. Usmanov, G. Zank and A. Dosch for fruitful discussions. Financial support for the project FI 706/8-2 (within Research Unit 1048), as well as for the projects FI 706/14-1 and HE 3279/15-1 funded by the Deutsche Forschungsgemeinschaft (DFG), and FWF-Projekt I1111
is acknowledged. 





\newpage
\appendix
\section{Turbulence transport equations}
\label{appendixA0}
The turbulence transport equations used in \citet{Usmanov-etal-2011} can be derived from the more general model of \citet{Zank-etal-2012a} with the simplifying assumptions of a single correlation length $\lambda=\lambda^\pm=\lambda_D/2$ and a choice of the structural similarity parameters $a=1/2, b=0$ corresponding to axisymmetric turbulence along the mean magnetic field direction, i.e. $\bf{\hat{n}}=\bf{\hat{B}}$. In not neglecting the Alfv\'{e}n velocity and considering the co-rotating frame of reference we get from equations (37), (38) and (35) of \citet{Zank-etal-2012a}:
 \begin{align}
\partial_t Z^2 &+ {\bf V}\cdot\nabla Z^2 + \frac{Z^2}{2}\nabla\cdot{\bf U} - {\bf V}_A\cdot\nabla(Z^2\sigma_C) + Z^2\sigma_C\nabla\cdot{\bf V}_A \nonumber \\
&+ Z^2\sigma_D\left(\frac{\nabla\cdot{\bf U}}{2}- {\bf\hat{B}}\cdot({\bf\hat{B}}\cdot\nabla){\bf U}\right) \nonumber \\
&= - \frac{2Z^3f^+(\sigma_C)}{\lambda_{Zank}} + \langle{\bf z^+}\cdot{\bf S^+}\rangle +  \langle{\bf z^-}\cdot{\bf S^-}\rangle  \\
\partial_t (Z^2\sigma_C) &+ {\bf V}\cdot\nabla(Z^2\sigma_C) + \frac{Z^2\sigma_C}{2}\nabla\cdot{\bf U} - {\bf V}_A\cdot\nabla Z^2 + Z^2(1-\sigma_D)\nabla\cdot{\bf V}_A \nonumber \\
&= - \frac{2Z^3f^-(\sigma_C)}{\lambda_{Zank}} + \langle{\bf z^+}\cdot{\bf S^+}\rangle -  \langle{\bf z^-}\cdot{\bf S^-}\rangle  \\
\partial_t\lambda_{Zank} &+ {\bf V}\cdot\nabla\lambda_{Zank}  \nonumber \\
&= \frac{2Zf^+(\sigma_C)}{\sqrt{1-\sigma_C^2}} - \frac{\lambda_{Zank}}{Z^2}\left(\frac{\langle{\bf z^+}\cdot{\bf S^+}\rangle(1-\sigma_C) + \langle{\bf z^-}\cdot{\bf S^-}\rangle(1+\sigma_C)}{1-\sigma_C^2} \right)~, \label{eq:lambda_app}
\end{align}
with $f^\pm = \sqrt{1-\sigma_C^2}\left[\sqrt{1+\sigma_C}\pm\sqrt{1-\sigma_C}\right]$
where the following moments of the Els\"asser variables $\bf{z^\pm} := \bf{u} \pm \bf{b}/\sqrt{\rho}$ ($\bf{u}$ and $\bf{b}$ denoting the fluctuations about the mean fields $\bf{U}$ and $\bf{B}$) are used:\\
\begin{align}
Z^2 &:= \frac{\langle\bf{z^+}\cdot\bf{z^+}\rangle +  \langle\bf{z^-}\cdot\bf{z^-}\rangle}{2} = \langle u^2 \rangle + \langle b^2/\rho \rangle \ (=E_T) \\
Z^2\sigma_C &:= \frac{\langle\bf{z^+}\cdot\bf{z^+}\rangle -  \langle\bf{z^-}\cdot\bf{z^-}\rangle}{2} = 2\langle{\bf u}\cdot{\bf b}/\sqrt{\rho}\rangle \ (=E_C) \\
Z^2\sigma_D &:= \langle{\bf z^+}\cdot{\bf z^-}\rangle = \langle u^2 \rangle - \langle b^2/\rho \rangle \ (=E_D)~,
\end{align}
where the last equality is to clarify the change of notation from \citet{Zank-etal-2012a}. These authors absorbed a factor of 2 into the definition of the correlation length ($\lambda_{Zank}=2\lambda_{def}$). To make use of the conservative scheme in the {\sc Cronos} code we rewrite the above equations in the fashion of $\partial_t X + \nabla\cdot{\bf F}_X = S_X$. Removing the factor 2 in the correlation length we obtain:
\begin{align}
\partial_t Z^2 + \nabla\cdot({\bf U}Z^2 + {\bf V}_AZ^2\sigma_C) &= \frac{Z^2(1-\sigma_D)}{2}\nabla\cdot{\bf U} + 2{\bf V}_A\cdot\nabla(Z^2\sigma_C) + Z^2\sigma_D{\bf\hat{B}}\cdot({\bf\hat{B}}\cdot\nabla){\bf U} \nonumber \\
&- \frac{\alpha Z^3f^+(\sigma_C)}{\lambda} + \langle{\bf z^+}\cdot{\bf S^+}\rangle +  \langle{\bf z^-}\cdot{\bf S^-}\rangle \\
\partial_t(Z^2\sigma_C) + \nabla\cdot({\bf U}Z^2\sigma_C + {\bf V}_AZ^2) &= \frac{Z^2\sigma_C}{2}\nabla\cdot{\bf U} + 2{\bf V}_A\cdot\nabla Z^2 + Z^2\sigma_D\nabla\cdot{\bf V}_A \nonumber \\ 
&- \frac{\alpha Z^3f^-(\sigma_C)}{\lambda} + \langle{\bf z^+}\cdot{\bf S^+}\rangle - \langle{\bf z^-}\cdot{\bf S^-}\rangle \\
\partial_t(\rho\lambda) + \nabla\cdot({\bf U}\rho\lambda) &= \rho\beta\left[Zf^+(\sigma_C) - \frac{\lambda}{\alpha Z^2}\left(\langle{\bf z^+}\cdot{\bf S^+}\rangle(1-\sigma_C) + \langle{\bf z^-}\cdot{\bf S^-}\rangle(1+\sigma_C) \right)\right] \nonumber \\ 
\end{align}
Here, we have also made the following adaptions in order to obtain the slightly different dissipation terms used by \citet{Usmanov-etal-2011}, whose model we use to validate our implementation:
Involving the Karman-Taylor constants $\alpha=2\beta=0.8$ in the dissipation terms and neglecting additional factors of $1-\sigma_C^2$ on the right hand side of equation (\ref{eq:lambda_app}) (due to different modeling of this evolution equation, here it is taken from \citet{Breech-etal-2008}).\\ 
On neglecting Alfv\'{e}n velocity terms and using only the isotropization of newly born pickup ions as source for turbulence, i.e.~$\langle{\bf z^\pm}\cdot{\bf S^\pm}\rangle = \dot{E}_{pui}/2 = 0.5f_dUV_An_H/(n_0\tau_{ion})\exp(-L_{cav}/r)$, we arrive at the respective equations used in \citet{Usmanov-etal-2011}.

\section{Energy equation}
\label{appendixB}
The total energydensity to be conserved in the coupled MHD -- turbulence transport model is
\begin{equation}
E = \rho U^2/2 + p/(\gamma-1) + B^2/2 - \rho GM_\odot/r + \rho Z^2/2 \label{eq:totE}
\end{equation}
accounting for the (rest-frame) kinetic, thermal, magnetic, gravitational, and turbulent energy densities. The resulting conservation equation can be obtained from carrying out a time derivative on (\ref{eq:totE}), which gives after some algebra \citep[see Appendix A of][]{Usmanov-etal-2011}
\begin{equation}
\partial_t E + \nabla\cdot\left[{\bf V}E+{\bf U}\bar{p}-\eta({\bf U} \cdot {\bf B}) {\bf B} + {\bf q}_H\right] = (\langle{\bf z^+}\cdot{\bf S^+}\rangle +  \langle{\bf z^-}\cdot{\bf S^-}\rangle)\rho/2 ~.
\label{eq:E}
\end{equation}
with $\eta = 1 + \sigma_D\rho Z^2/(2B^2)$, $\bar{p} = p + B^2/2 + p_w$ and $p_w = (\sigma_D+1)\rho Z^2/4$.
The current version of the {\sc Cronos} code does not provide the possibility to incorporate additional forms of energy densities with the original 
$$e = \rho U^2/2 + p/(\gamma-1) + B^2/2$$
obtained for the case of ideal MHD without any source terms. Therefore, in order to maintain a conservation of all involved forms of energies, we have to introduce respective source terms $Q$ in the conservation equation of $e$ 
\begin{equation}
	\partial_t e + \nabla\cdot\left[{\bf V}e+{\bf U}(p+B^2/2)-({\bf U} \cdot {\bf B}) {\bf B} + {\bf q}_H\right] = Q
\end{equation}
such that on substracting equation (\ref{eq:E}) and solving for $Q$ we obtain
\begin{align}
	Q &= -\partial_t(\rho Z^2/2) - \nabla\cdot[{\bf V}(\rho Z^2/2 - \rho GM_\odot/r)] \nonumber \\
	&- \nabla\cdot[{\bf U}p_w - ({\bf U} \cdot {\bf B}) {\bf B}\sigma_D\rho Z^2/(2B^2)] + (\langle{\bf z^+}\cdot{\bf S^+}\rangle +  \langle{\bf z^-}\cdot{\bf S^-}\rangle)\rho/2 ~.
\end{align}
After some lengthy algebra and involving equations (\ref{eq:conti}) and (\ref{eq:Z2}) this gives
\begin{align}
	Q &= \left[Z^2\sigma_C\nabla\cdot{\bf V}_A-{\bf V}_A\cdot\nabla(Z^2\sigma_C)\right]\frac{\rho}{2} - {\bf U}\cdot\nabla p_w + {\bf U}\cdot({\bf B}\cdot\nabla)[(\eta-1){\bf B}] \nonumber \\
	&+ \frac{\rho Z^3f^+(\sigma_C)}{2\lambda} - \rho{\bf V}\cdot{\bf g}
\end{align}
where we also used the vector identity 
\begin{equation}
{\bf U}\cdot({\bf B}\cdot\nabla)[(\eta-1){\bf B}] = \nabla\cdot(\eta-1){\bf B}({\bf U}\cdot{\bf B}) - (\eta-1){\bf B}\cdot({\bf B}\cdot\nabla){\bf U} ~.
\end{equation}
Finally, on extracting a flux term viz. 
\begin{equation}
\left[Z^2\sigma_C\nabla\cdot{\bf V}_A-{\bf V}_A\cdot\nabla(Z^2\sigma_C)\right]\frac{\rho}{2} = \nabla\cdot(Z^2\sigma_C{\bf V}_A\rho/2) - \rho{\bf V}_A\cdot\nabla(Z^2\sigma_C) - Z^2\sigma_C{\bf V}_A\cdot\nabla\rho/2
\end{equation}
and adding Hollweg's heat flux \citep{Hollweg-1974, Hollweg-1976} we get the final equation
\begin{align}
\partial_t e &+ \nabla \cdot \left[ e{\bf V} + (p+ |{\bf B}|^2/2) \ {\bf U}
    - ({\bf U} \cdot {\bf B}) {\bf B} - {\bf V}_A\rho Z^2\sigma_C/2 + {\bf q}_H \right] = \nonumber \\
		&-\rho{\bf V}\cdot{\bf g} -{\bf U}\cdot\nabla p_w - \frac{Z^2\sigma_C}{2}{\bf V}_A\cdot\nabla\rho + \frac{\rho Z^3f^+(\sigma_C)}{2\lambda}\nonumber \\
	&+ {\bf U}\cdot({\bf B}\cdot\nabla)[(\eta-1){\bf B}] - \rho{\bf V}_A\cdot\nabla(Z^2\sigma_C) ~. 
\end{align}

\section{The CRONOS code}
\label{app_cronos}
The {\sc Cronos} code used in this study is a versatile code for the numerical solution of the MHD equations. The code is written in C++ and is fully MPI-parallel. Usage of C++ leads to a high modularity that allows changing core components at runtime or adding new features rather easily.\\
The code is of second order in space and time. Among the core components are a variety of approximate Riemann solvers that can be chosen by the user according to the model to be simulated. {\sc Cronos} solves hydrodynamical (HD) and magnetohydrodynamical (MHD) problems. The corresponding Riemann solvers included in {\sc Cronos} are {\sc Hll} \citep[see][]{Harten-etal-1983}, {\sc Hllc} \citep[HD only, see][]{Toro-etal-1994} and {\sc Hlld} \citep[MHD only, see][]{Miyoshi-Kusano-2005}.\\
For MHD simulations the solenoidality of the magnetic field is ensured by using constrained transport \citep[see][who first applied it to MHD]{Brackbill-Barnes-1980}. For the {\sc Hll} solver the numerical electric field computation is consistently implemented as described in \citet{Kissmann-Pomoell-2012} \citep[see also in][]{Londrillo-delZanna-2000,Londrillo-delZanna-2004,Ziegler-2011}. For the {\sc Hlld} we implemented the electric field computation as described in \citet{Gardiner-Stone-2005}, who showed that a computation via a direct averaging of the fluxes resulting from the conservative form of the MHD equations can lead to instabilities.\\
Since {\sc Cronos} is optimised for highly compressible flows, the second-order reconstruction employs slope limiters that can be chosen by the user. Other limiters can be added easily to the simulation framework. Numerical problems are solved on an orthogonal grid, where Cartesian, cylindrical, and spherical coordinates are supported. For such a grid the cell size can be varied along each coordinated direction. The grid is specified by the user and is not changed during the simulation.\\
The code is used by supplying a simulation module that contains all relevant information for the simulation setup. Apart from the initial conditions, e.g.~additional forces can be defined by the user. A feature extensively used in the present study is the option to solve other hyperbolic partial differential equations simultaneously with the system of MHD equations. For this, the specific fluxes of the additional differential equations can also be specified within the user module. In a similar fashion, other types of differential equations can be handled in parallel to the main solver by supplying the corresponding alternative solver. {\sc Cronos} offers a standard interface to specify the alternative solver. For the solution in parallel to the main solver an operator-splitting approach is implemented. \\
The {\sc Cronos} code has been applied in a range of different studies ranging from stellar wind simulations to accretion disc models \citep[see][]{Flaig-etal-2011}. Verification of the code was done in the context of the different studies. In particular the numerical papers \citet{Kissmann-etal-2009} and \citet{Kissmann-Pomoell-2012} show that various HD and MHD standard problems are solved correctly by the code.




\clearpage

\end{document}